# A Density Functional Study of $O_2$ Adsorption on (100) Surface of γ-Uranium

M. N. Huda and A. K. Ray*
*P.O. Box 19059, Department of Physics, The University of Texas at Arlington*
Arlington, Texas-76019

We have studied the chemisorption processes of $O_2$ on the (100) surface of uranium using generalized gradient approximation to density functional theory. Dissociative adsorptions of $O_2$ are significantly favored compared to molecular adsorptions. We found interstitial adsorption of molecular oxygen to be less probable, as no bound states were found in this case. Upon oxygen adsorption, O 2p orbitals is found to hybridize with U 5f bands, and part of the U 5f electrons become more localized. Also there is a significant charge transfer from the first layer of the uranium surface to the oxygen atoms, which made the bonding partly ionic. For the most favored site the dissociative chemisorption energy is approximately 9.5 eV, which indicates a strong reaction of uranium surface with oxygen. Spin polarization does not have a considerable effect on the chemisorption process. For most of the sites and approaches, chemisorption configurations are almost same at both spin-polarized and non-spin-polarized cases. For the most favored chemisorption sites of oxygen on uranium, paramagnetic adsorption is slightly stronger, by 0.304 eV, than the magnetic adsorption.

**A. Introduction:**

Considerable theoretical efforts have been devoted in recent years to studying the electronic and geometric structures and related properties of surfaces to high accuracy. One of the many motivations for this burgeoning effort has been a desire to understand the detailed mechanisms that lead to surface corrosion in the presence of environmental gases; a problem that is not only scientifically and technologically challenging but also environmentally important. Such efforts are particularly important for systems like the actinides for which experimental work is relatively difficult to perform due to material problems and toxicity. As is known, the actinides are characterized by a gradual filling of the 5f-electron shell with the degree of localization increasing with the atomic number Z along the last series of the periodic table. The open shell of the 5f electrons determines the magnetic and solid state properties of the actinide elements and their compounds and understanding the quantum mechanics of the 5f electrons is the defining issue in the physics and chemistry of the actinide elements. These elements are also characterized by the increasing prominence of relativistic effects and their studies can, in fact, help us understand the role of relativity throughout the periodic table. Narrower 5*f* bands near the Fermi level, compared to 4*d* and 5*d* bands in transition elements, is believed to be responsible for the exotic structure of actinides at ambient condition [1]. The 5f orbitals have properties intermediate between those of localized 4f and delocalized 3d orbitals and as such, the actinides constitute the "missing link" between the d transition elements and the lanthanides [2]. Thus a proper and accurate understanding of the actinides will help us understand the behavior of the lanthanides and transition metals as well.

---
*email: akr@exchange.uta.edu

Uranium (U) is well known due its use as nuclear reactor fuel and is the heaviest naturally occurring actinide element. It is located in the middle of the early part of the actinide series, with only three 5f electrons hybridizing with the 6d and 7s electrons and demonstrating itinerant behavior. The proportion of the outer shell s and d electrons is larger in uranium compared to plutonium and a study of the electronic structure of U can provide significant clues about the crossover from delocalized to localized 5f-electron behavior [3]. Uranium crystallizes in the orthorhombic α-phase with four molecules per unit cell at ambient condition, followed by the body-centered tetragonal β (bct) phase at 940 K and then the γ(bcc) phase at 1050 K at ambient pressure. However, certain impurities like molybdenum can stabilize the γ-phase at room temperature or below [4]. The unfilled narrow 5*f* bands and the complexities in bonding in U might arise from the fact that it has valence shell which breaks Hund's third rule [5]. U also was one of the first examples of metal that undergoes superconducting transition under pressure without crystallographic transition [6]. These unusual aspects of the electronic bonding and structures in bulk uranium are apt to be enhanced at a surface or in a thin layer of uranium adsorbed on a substrate, due to the reduced atomic coordination of a surface atom and the narrow bandwidth of the surface states. For this reason, uranium surfaces and films and adsorptions on such may provide a valuable source of information about the bonding in uranium. Hao *et al.* [3] in a study of a five layers slab of (100) γ-uranium surface using the film-linearized-muffin-tin-orbitals (FLMTO) method suggested that surface enhancement of 5f localization (relative to bulk) is much stronger for uranium than for plutonium, with important consequences for surface reconstruction, chemisorption, and other surface behavior. Similar claims about γ-U having some localized 5*f* electrons have also been made in the literature [7]. In contrast, using surface spectroscopic techniques such as XPS, UPS, and AES, Gouder [8] concluded that the localization effects are strong in Pu films, whereas in U films effects are weak. Considering the narrow bandwidth of surface states, any transition from itinerant to localized behavior probably first takes place at the U surface with possible relaxations and reconstructions.

The uranium-oxygen system is one of the most complex metal oxide systems due to the high reactivity of U with $O_2$ and towards oxygen containing systems such as $H_2O$, CO and $CO_2$. A large number of oxide phases exist with a wide variety of stoichiometry [9]. Oxidation of metallic uranium surface has its technological importance primarily because of the atmospheric corrosion of uranium, and the formation of passivation layers protecting further corrosion attack [10]. At temperature below 35$^o$C, the reaction of uranium with water is totally suppressed and U+$O_2$ becomes the preferred reactions [11]. McLean *et al.* [12] used x-ray photoelectron spectroscopy, Auger electron spectroscopy and second ion mass spectroscopy to study $O_2$, CO and $CO_2$ on thorium and uranium surface. They showed that the adsorbed molecules dissociate and the carbon defused in the bulk, whereas the oxygen remained on the surfaces as an oxide. They also mentioned that the spectrum of uranium at saturation oxygen coverage closely resembles to that of $UO_2$. A similar study by Bloch *et al.* [10] reached the same conclusion. On the study of the progression of U-O surface reaction, this group later showed that the chemisorbed oxygen formed islands on the uranium surface, later spreading over the surface. Gouder *et al.* [13] used ultraviolet photospectroscopy to study the reaction of $O_2$ on uranium surface and showed that dissociative chemisorption of oxygen is followed by the formation of substoichiometric $UO_{2-x}$ and hyperstoichiometric $UO_{2+x}$ on the surface. They also found that $O_2$ adsorption results in a decrease of Fermi level emission and the increase of the U $5f^2$ and O 2p emission, which means the withdrawal of the 5f electrons from the Fermi level and their transfer into O 2p and the localized U $5f^2$ level.

On the theoretical side, there are *no* results in the literature on molecular oxygen chemisorption, including preferred adsorption sites and chemisorption energies on uranium surfaces. However, using the linear combination of Gaussian type orbitals – fitting function (LCGTO-FF) method within the GGA approximation of density functional theory (GGA-DFT), Boettger and Ray have carried out detailed electronic structure studies of crystalline $UO_2$ and its magnetic ordering [14]. Hybrid density functional theory with relativistic effective core potentials (RECP) has been used by Kudin *et al.* [15] to study the insulating gap of $UO_2$. The purpose of this paper is to present detailed electronic and geometric structure studies of the initial stages of oxygen chemisorption process on the uranium (100) surface by using *ab initio* methods. We first discuss the computational formalism followed by the results.

**B. Computational Details and Results**

As in our previous works [16], all computations reported here have been performed at the spin restricted and unrestricted generalized gradient approximation (GGA) level to [17] density functional theory (DFT) [18], using the suite of programs DMol3 [19]. In DMol3, the physical wave function is expanded in accurate numerical basis set and fast convergent three-dimensional integration is used to calculate the matrix elements occurring in the Ritz variational method. For the H atom, a double numerical basis set with polarization functions (DNP) and a real space cut-off of 5.0 A was used. The sizes of these DNP basis set are comparable to the 6-31G** basis of Hehre *et al.* [20]. However, they are believed to be much more accurate than a Gaussian basis set of the same size [19]. For U, the outer fourteen electrons ($6s^2\ 6p^6\ 5f^3\ 6d^1\ 7s^2$) are treated as valence electrons and the remaining seventy-eight electrons are treated as core. A hardness conserving semi-local pseudopotential, called density functional semi-core pseudo-potential (DSPP) [19], has been used. These norm-conserving pseudo-potentials are generated by fitting all-electron relativistic DFT results and have a non-local contribution for each channel up to $l = 2$, as well as a non-local contribution to account for higher channels. To simulate periodic boundary conditions, a vacuum layer of 30 Å was added to the unit cell of the layers. The k-point sampling was done by the use of Monkhorst-Pack scheme [21]. The maximum number of numerical integration mesh points available in DMol3 has been chosen for our computations, and the threshold of density matrix convergence is set to $10^{-6}$.

Though the uranium metal is believed to be paramagnetic, an ultra thin film (UTF) of uranium could be magnetic due to the local magnetic ordering at the narrower electronic bands on the surface. Also, it is well known that $UO_2$ becomes a noncollinear antiferromagnet below 30.8 K [22]. Thus, to understand the influence of magnetism on the chemisorption process, we have performed both non-spin-polarized and spin-polarized calculations. . As for the effects of relativity are concerned, DMol3 does not yet allow fully relativistic computations and as such, we have used the scalar-relativistic approach, as available in Dmol3. In this approach, the effects of spin-orbit coupling is omitted primarily for computational reasons but all other relativistic kinematic effects such as mass-velocity, Darwin, and higher order terms are retained. It has been shown [19] that this approach models actinide bond lengths fairly well. We certainly do not expect that the inclusion of the effects of spin-orbit coupling, though desirable, will alter the primary qualitative and quantitative conclusions of this paper, particularly since we are interested in chemisorption energies defined as the difference in total energies. Boettger and Ray [14] noted in their uranium dioxide study that the spin-polarized induced splitting of U 5f bands is roughly 1.0eV, compared to the spin-orbit splitting of 0.3eV. Hay and Martin [23] found that one could adequately describe the electronic and geometric properties of actinide complexes without treating spin-orbit effects explicitly. Similar conclusions have been reached by Ismail *et al*. [24] in their study of uranyl and plutonyl ions. We also note, as mentioned before, that scalar-relativistic hybrid density functional theory has been used by Kudin *et al*. [15] to describe the insulating gap of $UO_2$, yielding a correct anti-ferromagnetic insulator. All calculations are done on a Compaq ES40 alpha multi-processor supercomputer at the University of Texas at Arlington.

To study $O_2$ adsorption on the γ-U (100) surface, we have modeled the surface with three layers of uranium at the experimental lattice constant. This is believed to be quite adequate considering that the oxygen molecule is not expected to interact with atoms beyond the first three layers. This has been found to be the case in our studies of oxygen and hydrogen atom adsorptions on the plutonium surface [25]. Recently, Ray and Boettger [26] showed in a study of quantum size effects of δ-plutonium surface that surface energies converge within the first three layers. Due to severe demands on computational resources, the unit cell per layer was chosen to contain four U atoms. Thus our three-layer model of the surface contains twelve U atoms. The $O_2$ molecule, one per unit cell, was allowed to approach the uranium surface along three different symmetrical sites: i) directly on top of a U atom (*top* site); ii) on the middle of two nearest neighbor U atoms (*bridge* site); iii) in the center of the smallest unit structures of the surfaces (*center* site). As the smallest structure of (100) γ-uranium surface is a square, these three sites are the only symmetrically distinguishable sites. In addition to this, we have also considered some positions inside the U three layers slab (*interstitial* positions). Again for each of these positions, we consider several approaches of chemisorptions. They are: *a*) $O_2$ molecule approaches vertically to the surface (Ver approach), *b*) $O_2$ parallel to the surface and parallel to the bcc lattice vectors (Hor1 approach), and *c*) $O_2$ parallel to the surface and have an angle $45^o$ with the bcc lattice vectors, *i.e.*, parallel to the diagonal of the square lattice (Hor2 approach). It is obvious that for both of the horizontal approaches the oxygen atoms of $O_2$ are at the same distance from the uranium surface, whereas for the vertical approach one oxygen atom is closer to the surface than the other. For geometry optimization, the distances of two oxygen atoms from the surface ($r_d$) and the distance between the atoms ($r_O$) are both simultaneously optimized. The chemisorption energies are then calculated from:

$$E_c = E \text{ (U-layers)} + E \text{ } (O_2) - E \text{ (U-layers} + O_2) \qquad (1)$$

For the non-spin-polarized case, both E (U-layers) and E (U-layers + $O_2$) were calculated without spin polarization, while for spin polarized chemisorption energies both of these energies are spin polarized. $E(O_2)$ is the energy of the oxygen molecule in its triplet state in both of the cases. The chemisorption energies, and the corresponding distances are given in table 1. The distances $r_d$ given in the table are measured as the distance from the uranium surface to the oxygen atoms, if both the oxygen atoms are at same height, or to the nearer oxygen atom if one of them is closer to the surface than the other.

We start by describing the chemisorption process of $O_2$ at the different sites on uranium surfaces. Consider first the top sites without spin polarization (figure 1). It was mentioned earlier that there are three different approaches for each sites. For the two horizontal approaches the chemisorption distances ($r_d$) from the uranium surface to the $O_2$ are 2.33 Å and 1.99 Å; while the chemisprption energies are, 9.05 eV and 3.56 eV for *Hor1* and *Hor2*, respectively. In the Hor1 approach, the distance between the oxygen atoms is 3.47 Å, which imply complete dissociation of $O_2$. On the other hand in Hor2 approach the $O_2$ bond length stretched up to 1.51 Å. The O-O distance for the vertical approach is the lowest with the least $O_2$ adsorption energy, which is 1.96 eV. Inclusion of spin polarization did not change the over all feature of Ver and Hor2 approaches, while for Hor1 approach $O_2$ did not dissociate, as opposed to non-spin polarized case, with lower chemisorption energy.

For the bridge sites, the chemisorptions of $O_2$ along the vertical approach behaved differently for non-spin-polarized and spin-polarized cases. For the non-spin-polarized case, $O_2$ remained as a molecule, while for spin-polarized case the oxygen molecule dissociate and the final adsorption sites resembles to the non-spin-polarized case of top site at Hor1 approach. The O-O relaxation for this approach is 3.47 Å, and the chemisorption energy is 8.89 eV. For the *Hor1* approach at bridge sites, $O_2$ completely dissociates, and the inclusion of spin orbit coupling changes the geometry of final chemisorption sites (see figure 2), with slightly higher chemisorption energy. The spin polarization increased the $r_O$ and $r_d$ distances by 0.37 Å and 0.16 Å, respectively, from the non-spin-polarized case. For the Hor2 approach the oxygen molecule dissociate, but the spin polarization does not have any considerable effect on chemisorption geometry. However the non-spin-polarized energy is 0.117 eV higher than its spin-polarized counterpart. In general, chemisorption at the bridge site is considerably stronger than at the top site. This results from fact that oxygen atoms are relatively much closer to the uranium surface in bridge sites compared to the top sites. However, we note that non-spin polarized cases of Hor1 approach between the two sites behaved differently. For the bridge site at Hor1 approach, oxygen atoms are closer to the Pu surface ($r_d$ = 0.68 Å) than that of the top site ($r_d$ = 2.33 Å). However, the chemisorption energy is lower. The reason might be that the nearest U-O distances in the top site are smaller than the bridge site. In the top site the nearest U-O distances are 2.12 Å and each oxygen atom is two-fold coordinated; while for the bridge the corresponding nearest distance is 2.41 Å and is one-fold coordinated. Another point to be noted that in this case for the top site O-O distances is larger than that of bridge site. Larger O-O distance means less coulomb repulsion force, as both of the adsorbed oxygen atoms usually get negatively charged.

For the center sites at Ver approach, like bridge sites, in spin polarized case $O_2$ dissociated completely with O-O distance of 4.84 Å, while for the non-spin polarized case $r_O$ is 1.48 Å. Consequently the chemisorption energy is much higher, 9.154 eV, for the spin-polarized case compared to the corresponding non-spin-polarized case, 2.61 eV. For Hor1 approach, where after dissociation the oxygen atoms sit exactly in between two nearest neighbor uranium atoms (figure 3), the chemisorption energy, 9.492 eV, is the highest among all the possible configurations studied here. Inclusion of spin polarization lowers this value to 9.188 eV. For the Hor2 sites the O-O relaxation is up to 4.90 Å, with chemisorption energy of 9.237 eV (4.77 Å, 9.113 eV with spin polarization). The adsorption distances, $r_d$, are minimum for the center site, and in general the chemisorption energies are higher than the other two sites, except for the Hor1 approach in center and bridge site where bridge site has lower $r_d$ than the center site. The reason is that for the bridge site, after dissociation each oxygen atom is in the center position whereas for center site after dissociation each oxygen atom moved to the site defined as bridge site. It is to be noted that the final optimized position of oxygen atoms on uranium surface for spin-polarized Ver approach and spin- and non-spin-polarized case of Hor2 approach for center site are almost the same, and similar to the Hor2 approach of top site if the $O_2$ dissociated.

From the above discussion, it is clear that vertical approaches where the $O_2$ adsorption is molecular, are not the preferred approaches at any sites for the chemisorption processes. They have significantly lower chemisorption energies compared to the other cases where $O_2$ dissociates. Basically in molecular adsorption at vertical approach, one oxygen atom closer to the uranium surface is coordinated with uranium surface atoms, while the other one is only bonded with oxygen atom. This explains the much lower chemisorption energies of the vertical approach despite of the fact that oxygen atom is much closer to the surface. For example, in the center sites at vertical approach in non-spin-polarized case, the first oxygen atom is at 0.70 Å from the uranium surface and the second one is at 2.18 Å. Consequently the charge transfer to the lower atom (-0.47$e$) is slightly larger than the higher atom (-0.44$e$). The surface basically interacts with the first atom, while the coordination with the second atom to the surface is screened by the first oxygen atom. Except for the top position, both the horizontal approaches give comparable results, with overall preference depending on the U-O coordination.

Table 2 lists the Mulliken charge distribution [27] for the most stable chemisorption site in the spin-polarized case, namely the center site at the *Hor1* approach. Both the oxygen atoms acquire negative charges, 0.633$e$ each, primarily from the first layer of uranium atoms, with the first layer being positively charged as a result. Thus, there exists an ionic part in the U-O bonding, along with other contributions. A close look at the Hor1 and Hor2 approaches of center sites shows that, at Hor2 approach $r_d$ is smaller and $r_O$ is compared to the values of the Hor1 case, and also the charge transfers to the oxygen atoms are larger in Hor2 case (0.70$e$) than the Hor1 case (0.63$e$). Both of these facts imply that attractive coulomb force on the oxygen atom towards the surface is larger in Hor2 case than that of at Hor1 approach. However the fact that the chemisorption energy is slightly higher at Hor1 approach compared to the Hor2 approach implies that ionic part of the bonding is not strongly dominant. Also it is interesting to see that, though in the oxygen adsorbed uranium layers the second layer charge distribution is slightly modified than the bare case, the third layer charge distribution is almost unchanged (table 2). So the effect of $O_2$ chemisorption on the third layer is negligible.

We also considered several interstitial positions and it was found none of them gave a bound state with or without spin polarization. In addition to this it was found that the spin polarization has considerable effect on the total energies as a function of distance of $O_2$ from the top of the surface for interstitial positions. However, the interstitial oxygen adsorption is possible after the pre-dissociative adsorption on the outer surface. For example, only after the complete dissociation of the $O_2$ molecule at the center site with Hor1 approach into its atomic components, the two oxygen atoms can then diffuse through the surface. The chemisorption energies then can be calculated with respect to the atomic oxygen atoms, as the diffusion is atomic, not molecular. Due to the higher chemisorption energies at the surface, oxygen atoms mainly form a layer on the uranium surface as an oxide, as is experimentally confirmed [12].

In table 3, we have tabulated the magnetic moments of the oxygen adsorbed uranium layers for different adsorption configurations. The magnetic moments of the bare uranium layers drop rapidly as the number of layers increase, from 4.345 Bohr magnetons ($\mu_B$) for the monolayer to 1.610 Bohr magnetons for the 3-layer. This indicates that the semi-infinite uranium metal surface might indeed be paramagnetic. However, we note that our value for the magnetic moment of three layers of uranium slab is higher than the spin magnetic moment of 0.84 $\mu_B$ per atom for α-uranium predicted by Stojic *et al.* [28] using the full-potential-linearized-augmented-plane-wave (FPLAPW) method in the generalized gradient approximation. Experimentally, it has been noted that oxygen adsorbed uranium behaved like the $UO_2$ surface [12]. We also note that though experimentally $UO_2$ is noncollinear antiferromagnet, density functional theory tends to predict it as a ferromagnet [14]. Though the oxygen adsorbed uranium surface configurations described above does not exactly correspond to the $UO_2$ surface, it is expected that the oxygen adsorbed uranium thin film would be magnetic. Most of the magnetic moments tabulated in table 3 are of very low value, and lack any specific orderings. Except for few strongly chemisorbed cases, most of the oxygen adsorbed layers are almost paramagnetic. The most favorable dissociative chemisorption configuration, the center site with Hor1 approach, has also one of the highest magnetic moment of 0.174 $\mu_B$ per atom. It should be noted that the non-spin-polarized chemisorption energies for this same configuration is slightly higher than that of the spin polarized value. This leads us to conclude that spin polarization does not play a strong role in the overall chemisorption process.

From the band energetics of the bare and oxygen adsorbed uranium layers, we also found that the change in band gaps due to the inclusion of spin polarization is very small. For example the energy gaps for uranium 6p and 5f bands without spin polarization is 14.806 eV compared to 14.436 eV is with spin polarization. For the oxygen adsorbed layers, (we will consider only the most favored chemisorption configuration, center site at Hor1 approach- dissociative adsorption) these values are 10.214 eV and 10.177 eV, respectively. It can be inferred from these energies that the adsorption of oxygen reduces the gaps. The main reason for this reduction is that the oxygen 2p orbitals hybridize with the lower end of uranium 5f orbitals and split the 5f band. There exist a small band gap of 2.204 eV (2.562 eV with spin polarization) between the hybridized O 2p- U 5f band and the remaining U 5f electrons. The width of the 5f band before oxygen adsorption is almost same as the unhybridized part of the oxygen adsorbed U 5f band, approximately 1.04 eV. Oxygen 2s electrons sit below the uranium 6p band and hybridize with it. For bare uranium, the top of 5f band is 1.011 eV (0.959 eV with spin polarization) below the Fermi level. After addition of oxygen the energy difference between the top of 5f and Fermi level remains almost the same as the bare uranium surface, namely 0.994 eV and 0.954 eV for non-spin and spin polarized case, respectively. It was found that the upon oxygen adsorption the lower part of the 5f band become more localized, where the upper part is always hybridized with the uranium 6d and 7s electrons.

In conclusion, we have studied the chemisorption processes of $O_2$ on the (100) surface of uranium using generalized gradient approximation to density functional theory. Dissociative adsorptions of $O_2$ are significantly favored compared to molecular adsorptions. We also found interstitial adsorption of molecular oxygen to be less probable, as no bound states were found in this case. Only after dissociation of $O_2$, atomic oxygen diffusion through the surface is possible. O 2p orbitals is found to hybridize with U 5f bands, and part of the U 5f electrons become more localized. Also there is a significant charge transfer from the first layer of the uranium surface to the oxygen atoms, which made the bonding partly ionic. For the most favored site the dissociative chemisorption energy is approximately 9.5 eV, which indicates a strong reaction of uranium surface with oxygen. Spin polarization does not have a considerable effect on the chemisorption process. For most of the sites and approaches, chemisorption configurations are almost same at both spin-polarized and non-spin-polarized cases. For the most favored chemisorption sites of oxygen on uranium, paramagnetic adsorption is slightly stronger, by 0.304 eV, than the magnetic adsorption.

This work is supported by the Chemical Sciences, Geosciences and Biosciences Division, Office of Basic Energy Sciences, Office of Science, U. S. Department of Energy (Grant No. DE-FG02-03ER15409) and the Welch Foundation, Houston, Texas (Grant No. Y-1525).

**References:**

[1] P. Söderlind, O. Eriksson, B. Johansson, J. M. Wills, and A. M. Boring, Nature **374**, 524 (1995).

[2] N. M. Edelstein, B. Johansson, and J. L. Smith, *"Magnetic and Solid State Properties,"* in *Opportunities and Challenges in Research with Transplutonium Elements,* National Academy Press, Washington, D. C. 299 (1983).

[3] B. R. Cooper, O. Eriksson, Y.-G. Hao, and G. W. Fernando, *"Surface Electronic Structure and Chemisorption of Plutonium and Uranium,"* in *Transuranium Elements: A Half Century*, edited by L. R. Morss and J. Fuger (American Chemical Society, Washington, D. C.) 365(1992) ; Y. G. Hao, O. Eriksson, G. W. Fernando, and B. R. Cooper, Phys. Rev. B, **47**, 6680 (1993).

[4] D. D. Koelling and A. J. Freeman, Phys. Rev. B **7**, 4454 (1973).

[5] *A*. Hjelm, O. Eriksson, and B. Johansson, Phys. Rev. Lett. **71**, 1459 (1993).

[6] J. C. Ho, N. E. Phillips, and T. F. Smith, Phys. Rev. Lett. **17**, 694 (1966).

[7] C.-S. Yoo, H. Cynn and P. Söderlind, Phys. Rev. B **57**, 10359 (1998); P. Soderlind, *ibid,* **66**, 085113 (2002).

[8] T. Gouder, J. Alloys. Comp. **271**, 841 (1998)

[9] C. A. Colmenares, Progr. Solid State Chem., **9**, 139 (1975).

[10] J. Bloch, U. Atzmony, M. P. Dariel, M. H. Mintz and N. Shamir, J. Nucl. Mat., **105**, 196 (1982); E. Swissa, J. Bloch, U. Atzmony, and M. H. Mintz, Surf. Sci., **214**, 323 (1989).

[11] J. M. Haschke, J. Alloys. Comp., **278**, 149 (1998).

[12] W. McLean, C. A. Colmenares, R. L. Smith, and G. A. Somorjai, Phys. Rev. B, **25**, 8 (1982).

[13] T. Gouder, C. Colmenares, J. R. Naegele, and J. Verbist, Surf. Sci., **235**, 280 (1989).

[14] J. C. Boettger and A. K. Ray, Int. J. Quant. Chem., **80**, 824 (2000); *ibid,* **90**, 1470 (2002); J. C. Boettger, Eur. Phys. J., **36**, 15 (2003)

[15] K. N. Kudin, G. E. Scuseria, and R. Martin, Phys. Rev. Lett., **89**, 266402 (2002).

[16] X. Wu, *Density Functional Theory Applied to d- and f-Electron Systems*, Ph. D. Dissertation, The University of Texas at Arlington (2001); X. Wu and A. K. Ray, Phys. Rev. B **65**, 085403 (2002); Physica B **301**, 359 (2001); Eur. Phys. J. B **19**, 345 (2001).

[17] J. P. Perdew in *Electronic Structure of Solids,* edited by Ziesche and H. Eschrig (Akademie Verlag, Berlin, 1991); J. P. Perdew, K. Burke, and Y. Wang, Phys. Rev. B **54**, 16533 (1996); J. P. Perdew, K. Burke, and M. Ernzerhof, Phys. Rev. Lett. **77**, 3865 (1996).

[18] P. Hohenberg and W. Kohn, Phys. Rev. B **136**, 864 (1964); W. Kohn and L. J. Sham, Phys. Rev. A **140**, 1133 (1965); S. B. Trickey (Ed.), *Density Functional Theory for Many Fermion Systems* (Academic, San Diego, 1990); R. M. Dreialer and E. K. U. Gross, *Density Functional Theory: An Approach to Quantum Many Body Problem* (Springer, Berlin, 1990); J. F. Dobson, G. Vignale, and M. P. Das (Eds.), *Electronic Density Functional Theory − Recent Progress and New Directions* (Plenum, New York, 1998).

[19] B. Delley, J. Chem. Phys. **92**, 508 (1990); Int. J. Quant. Chem. **69**, 423 (1998); J. Chem. Phys. **113**, 7756 (2000); Phys. Rev. B **65**, 085403 (2002); A. Kessi and B. Delley, Int. J. Quant. Chem. **68**, 135 (1998).

[20] W. J. Hehre, L. Radom, P. v. R. Schlyer, and J. A. Pople, *Ab Initio Molecular Orbital Theory* (Wiley, New York, 1986).

[21] H. J. Monkhorst and J. D. Pack, Phys. Rev. B **13**, 5188 (1976).

[22] K. Ikushima, S. Tsutsui, Y. Haga, H. Yasuoka, R. E. Walstedt, N. M. Masiki, A. Nakamura, S. Nasu, and Y. Onuki, Phys. Rev. B **63**, 104404 (2001).

[23] P. J. Hay and R. L. Martin, J. Chem. Phys. **109**, 3875 (1998).

[24] N. Ismail, J.-L. Heully, T. Saue, J.–P. Daudey, and C. J. Marsden, Chem. Phys. Lett. **300**, 296 (1999).

[25] M. N. Huda and A. K. Ray, submitted for publication.

[26] A. K. Ray and J. C. Boettger, submitted for publication.

[27] R. S. Mulliken, J. Chem. Phys. **23**, 1833 (1955); *ibid*, **23**, 1841 (1955); **23**, 2343 (1955).

[28] N. Stojic, J. W. Davenport, M. Komelj, and J. Gilman, Phys. Rev. B **68**, 094407 (2003).

Table 1. Chemisorption energies in eV for different sites and approaches with adsorption distances, $r_d$ (in Å) from the uranium surface and the O-O distances, $r_O$ (in Å), are shown. For all the approaches $r_d$ is calculated from the lower oxygen atom to the uranium surfaces.

| Sites | Approach | $r_d$ in Å | $r_O$ in Å | Chemisorption Energy in eV |
|---|---|---|---|---|
| | | No Spin Polarization | | |
| Top | Ver | 2.03 | 1.33 | 1.960 |
| | Hor1 | 2.33 | 3.47 | 9.047 |
| | Hor2 | 1.99 | 1.51 | 3.557 |
| Bridge | Ver | 1.31 | 1.41 | 2.555 |
| | Hor1 | 0.68 | 3.09 | 8.740 |
| | Hor2 | 0.72 | 3.69 | 9.072 |
| Center | Ver | 0.70 | 1.48 | 2.612 |
| | Hor1 | 1.20 | 3.45 | 9.492 |
| | Hor2 | 0.65 | 4.90 | 9.237 |
| | | With Spin Polarization | | |
| Top | Ver | 2.01 | 1.32 | 1.998 |
| | Hor1 | 1.96 | 1.52 | 3.341 |
| | Hor2 | 1.98 | 1.50 | 3.546 |
| Bridge | Ver | 1.20 | 3.47 | 8.889 |
| | Hor1 | 0.84 | 3.46 | 8.830 |
| | Hor2 | 0.72 | 3.71 | 8.955 |
| Center | Ver | 0.62 | 4.84 | 9.154 |
| | Hor1 | 1.18 | 3.47 | 9.188 |
| | Hor2 | 0.63 | 4.77 | 9.113 |

Table 2. Spin and charge distribution of bare uranium layers and the most favorable chemisorption configurations (center site at *Hor*1 approach) are shown. The table below only shows the spin polarized case.

|  | Uranium layers | | Uranium + Oxygen layers | |
|---|---|---|---|---|
|  | Charge | Spin | Charge | Spin |
| O atoms | × | × | -0.633 | 0.045 |
|  | × | × | -0.633 | -0.001 |
| 1$^{st}$ layer | -0.079 | -2.054 | 0.257 | -1.279 |
|  | -0.079 | -2.054 | 0.252 | -0.905 |
|  | -0.079 | -2.054 | 0.252 | -0.921 |
|  | -0.079 | -2.054 | 0.251 | -0.979 |
| 2$^{nd}$ layer | 0.158 | -0.721 | 0.157 | 0.067 |
|  | 0.158 | -0.721 | 0.156 | 0.067 |
|  | 0.158 | -0.721 | 0.143 | 0.072 |
|  | 0.158 | -0.721 | 0.143 | 0.072 |
| 3$^{rd}$ layer | -0.079 | -2.054 | -0.084 | 2.229 |
|  | -0.079 | -2.054 | -0.086 | 1.936 |
|  | -0.079 | -2.054 | -0.087 | 1.983 |
|  | -0.079 | -2.054 | -0.089 | -1.901 |

Table 3. Magnetic moments of the oxygen chemisorbed uranium layers in Bohr magnetons, $\mu_B$, per atom are shown.

| Sites | Approach | Magnetic Moments in $\mu_B$ per atom |
|---|---|---|
| Top | Ver | 0.138 |
|  | Hor1 | 0.159 |
|  | Hor2 | 0.176 |
| Bridge | Ver | 0.021 |
|  | Hor1 | 0.032 |
|  | Hor2 | 0.008 |
| Center | Ver | 0.049 |
|  | Hor1 | 0.174 |
|  | Hor2 | 0.034 |

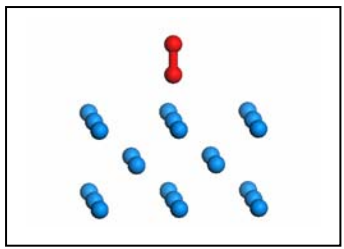

Top: Ver approach, NSP

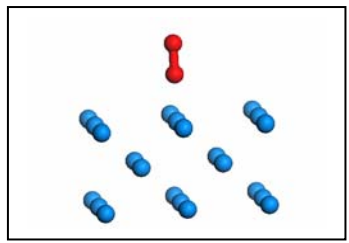

Top: Ver approach, SP

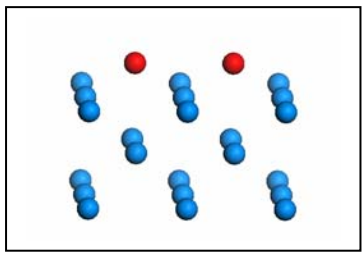

Top: Hor1 approach, NSP

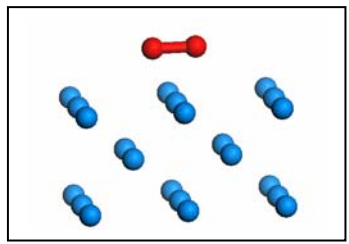

Top: Hor1 approach, SP

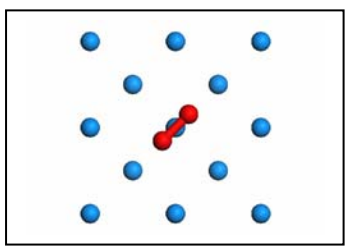

Top: Hor2 approach, NSP

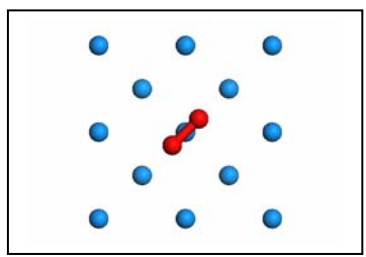

Top: Hor2 approach, SP

Figure 1. $O_2$ adsorption on uranium surface at top site.

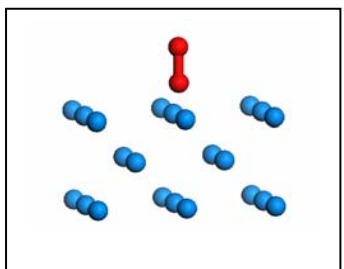
Bridge: Ver approach, NSP

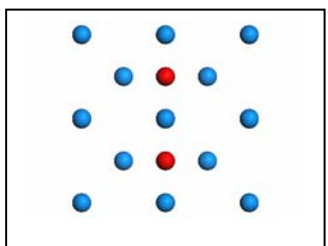
Bridge: Ver approach, SP

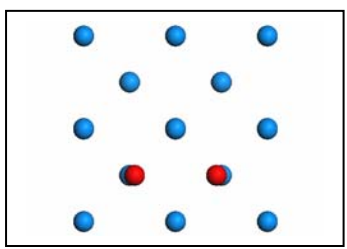
Bridge: Hor1 approach, NSP

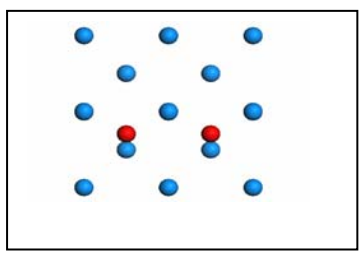
Bridge: Hor1 approach, SP

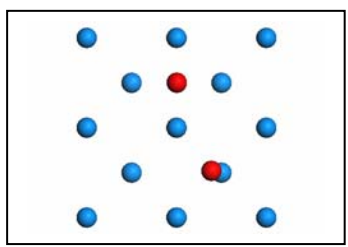
Bridge: Hor2 approach, NSP

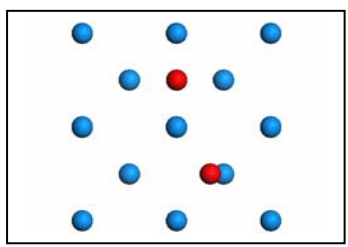
Bridge: Hor1 approach, SP

Figure 2. $O_2$ adsorption on uranium surface at bridge site.

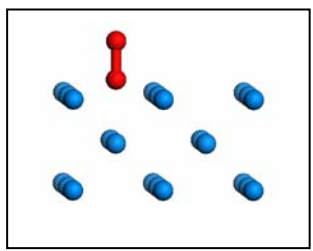 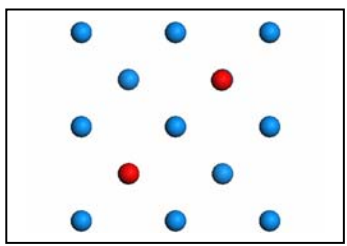

Center: Ver approach, NSP    Center: Ver approach, SP

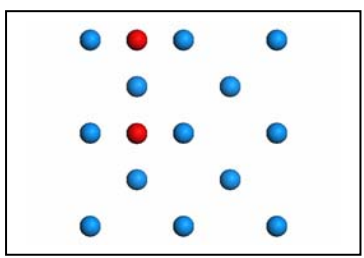 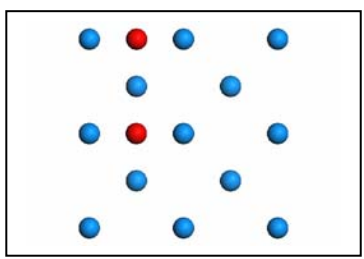

Center: Hor1 approach, NSP    Center: Hor1 approach, SP

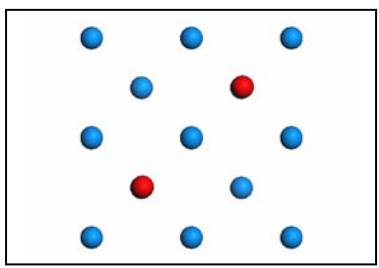 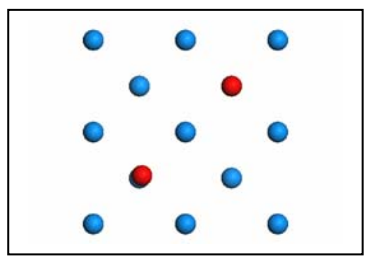

Center: Hor2 approach, NSP    Center: Hor2 approach, SP

Figure 3. $O_2$ adsorption on uranium surface at center site.